\documentclass[aps,twocolumn,preprintnumbers,amsmath,amssymb,prl]{revtex4}
\usepackage{graphicx}
\usepackage{bm}

\begin{document}

\title{Superconductivity induced by cobalt doping in iron-based oxyarsenides}

\author{Guanghan Cao\footnote[1]{Electronic address: ghcao@zju.edu.cn}, Cao Wang, Zengwei Zhu, Shuai Jiang, Yongkang Luo, Shun Chi, Zhi Ren, Qian Tao, Yuetao Wang and Zhu'an Xu\footnote[2]{Electronic address: zhuan@zju.edu.cn}}
\affiliation{Department of Physics, Zhejiang University, Hangzhou 310027, People's Republic of China}

\maketitle

\textbf{Chemical doping has recently become a very important strategy to induce superconductivity especially in
complex compounds. Distinguished examples include Ba-doped La$_2$CuO$_4$ (the first high temperature
superconductor)\cite{Bednorz&Muller}, K-doped BaBiO$_3$\cite{Cava1988}, K-doped C$_{60}$\cite{Hebard} and
Na$_{x}$CoO$_{2}\cdot y$H$_{2}$O\cite{Takada}. The most recent example is F-doped LaFeAsO\cite{Kamihara08},
which leads to a new class of high temperature superconductors. One notes that all the above dopants are
non-magnetic, because magnetic atoms generally break superconducting Cooper pairs. In addition, the doping site
was out of the (super)conducting structural unit (layer or framework). Here we report that superconductivity was
realized by doping magnetic element cobalt into the (super)conducting-active Fe$_2$As$_2$ layers in
LaFe$_{1-x}$Co$_{x}$AsO. At surprisingly small Co-doping level of $x$=0.025, the antiferromagnetic
spin-density-wave transition\cite{Dai} in the parent compound is completely suppressed, and superconductivity
with $T_c\sim $ 10 K emerges. With increasing Co content, $T_c$ shows a maximum of 13 K at $x\sim 0.075$, and
then drops to below 2 K at $x$=0.15. This result suggests essential differences between previous cuprate
superconductor and the present iron-based arsenide one.}
\\
\\
Recent discovery of superconductivity at 26 K in LaO$_{1-x}$F$_{x}$FeAs\cite{Kamihara08} has opened a new
chapter in superconductivity research. The superconductivity was induced by partial substitution of O$^{2-}$
with F$^{-}$ in the parent compound LaFeAsO whose crystal structure consists of insulating La$_2$O$_2$ layers
and conducting Fe$_2$As$_2$ layers (see the inset of Fig. 1(b)). Following this discovery, the superconducting
transition temperature $T_c$ over 40 K was realized in LnFeAsO$_{1-x}$F$_{x}$
(Ln=lanthanides)\cite{Chen-Sm,Chen-Ce,Ren-Sm} and LnFeAsO$_{1-\delta}$\cite{Ren-Sm2}. Through an alternative
chemical doping of thorium-for-gadolinium, $T_c$ has achieved 56 K in Gd$_{0.8}$Th$_{0.2}$FeAsO.\cite{Wang}
These substitutions introduce extra positive charges in the insulating Ln$_2$O$_2$ layers, and extra electrons
are produced onto the Fe$_2$As$_2$ layers as a result of charge neutrality. The occurrence of superconductivity
in this sense is rather similar to cuprate superconductors in which superconductivity appears when appropriate
amount of charge carriers are transferred into the CuO$_2$ planes by chemical doping at "charge reservoir
layers".\cite{Cava}

Band structure calculations and theoretical analysis reveal itinerant character of Fe 3$d$ electrons in the
iron-based oxyarsenides.\cite{Singh,wnl,Tesanovic} The calculated electron density-of-states (DOS) versus energy
for LaO$M$As ($M$ = Mn, Fe, Co and Ni) by Xu et al.\cite{Xu} shows that the main feature of total DOS remains
unchanged, except that Fermi levels shift toward the top of valence band with band filling (adding electrons)
one by one from $M$ = Mn, Fe, Co to Ni. According to this calculation, substitution of cobalt for iron is
expected to add electrons into Fe$_2$As$_2$ layers because cobalt has one more electron than iron does.
Therefore, we explored the cobalt substitution for iron in LaFeAsO system. Strikingly, superconductivity was
observed by slight cobalt doping even on the superconducting-active Fe$_2$As$_2$ layers.

The polycrystalline LaFe$_{1-x}$Co$_{x}$AsO samples were synthesized by solid state reaction in vacuum using
powders of LaAs, La$_{2}$O$_{3}$, FeAs, Fe$_{2}$As and Co$_{3}$O$_{4}$. LaAs was presynthesized by reacting
stoichiometric La pieces and As powders in evacuated quartz tubes at 1223 K for 24 hours. FeAs and Fe$_{2}$As
were prepared by reacting stoichiometric Fe powders and As powders at 873 K for 10 hours, respectively.
Co$_{3}$O$_{4}$ and La$_{2}$O$_{3}$ were dried by firing in air at 773 K and 1173 K, respectively, for 24 hours
before using. The powders of these intermediate materials were weighed according to the stoichiometric ratio of
LaFe$_{1-x}$Co$_{x}$AsO ($x$=0, 0.01, 0.025, 0.05, 0.075, 0.1, 0.125, 0.15 and 0.2) and thoroughly mixed in an
agate mortar and pressed into pellets under a pressure of 2000 kg/cm$^{3}$, all operating in a glove box filled
with high-purity argon. The pellets were sealed in evacuated quartz tubes and heated at 1433 K for 40 hours.

Figure 1(a) shows the representative XRD patterns of LaFe$_{1-x}$Co$_{x}$AsO samples. The XRD peaks can be well
indexed based on a tetragonal cell of ZrCuSiAs-type structure, indicating that the samples are essentially
single phase. The lattice parameters are plotted in Fig. 1(b) as a function of Co content. With increasing Co
content, the $a$-axis remains nearly unchanged while the $c$-axis shrinks significantly. Thus the cell volume
decreases almost linearly, which is related to the smaller Co$^{2+}$ ions (than Fe$^{2+}$ ions). This fact
indicates that Co was successfully doped into the lattice, according to Vegard's law.

\begin{figure}
\includegraphics[width=7cm]{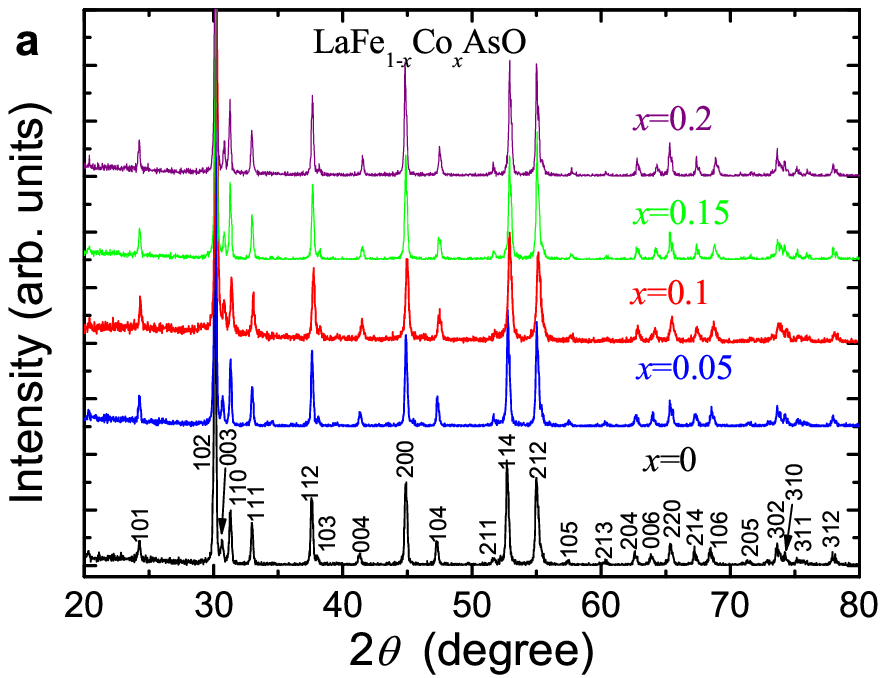}
\includegraphics[width=8cm]{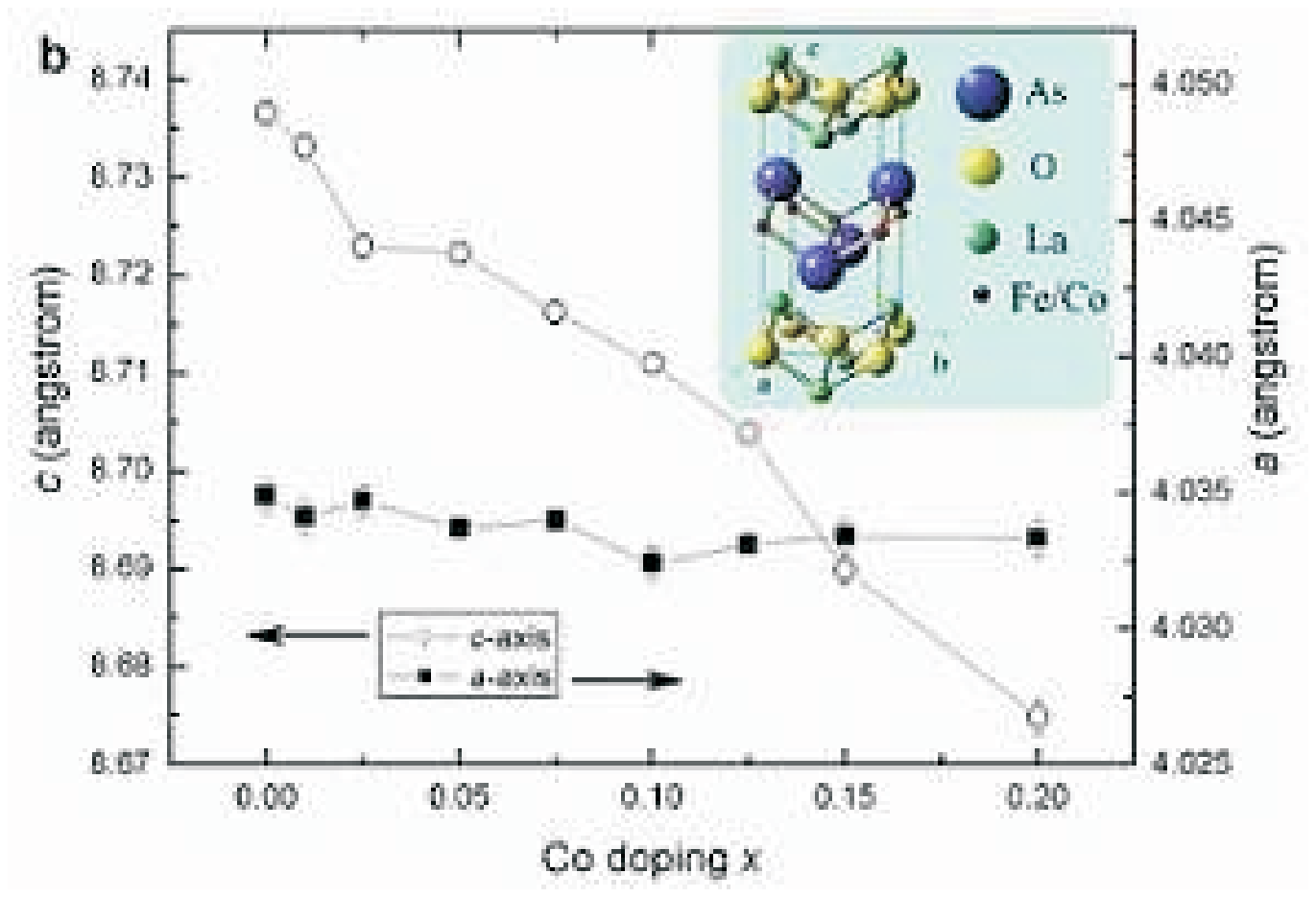}
\caption {\textbf{Structural characterization of LaFe$_{1-x}$Co$_{x}$AsO samples}. \textbf{a}, Powder X-ray
diffraction (XRD) patterns of LaFe$_{1-x}$Co$_{x}$AsO samples. XRD experiment was performed at room temperature
using a D/Max-rA diffractometer with Cu K$_{\alpha}$ radiation and a graphite monochromator. The diffractometer
system was calibrated using standard Si powders. \textbf{b}. Lattice parameters as a function of Co content. The
lattice parameters are refined based on a $P$4/$nmm$ space group by a least-squares fit using at least 20 XRD
peaks. The inset shows the crystal structure of LaFe$_{1-x}$Co$_{x}$AsO.}
\end{figure}

The oxygen content in LaFe$_{1-x}$Co$_{x}$AsO is an important issue in present study, because oxygen deficiency
itself might induce superconductivity. By high-pressure synthesis, superconductivity was indeed observed in
oxygen-deficient LnFeAsO$_{1-\delta}$\cite{Ren-Sm2,Kito}. It has also been reported that superconductivity was
induced by oxygen deficiency in Sr-doped LaFeAsO via annealing in vaccum.\cite{Wu} We note that all the reported
superconductors showed a remarkable decrease in $a$-axis as well as $c$-axis owing to the oxygen deficiency.
However, the present LaFe$_{1-x}$Co$_{x}$AsO samples show no obvious change in $a$-axis, suggesting no
significant oxygen deficiency.

Figure 2(a) shows the temperature dependence of electrical resistivity ($\rho$) in LaFe$_{1-x}$Co$_{x}$AsO. For
the parent compound, a prominent anomaly characterized by a drop of $\rho$ was observed below 150 K. Neutron
diffraction study\cite{Dai} indicated a structural phase transition at 155 K followed by an antiferromagnetic
spin-density-wave transition at 137 K in LaFeAsO. The drop in $\rho$ (and also $\chi$, shown in the inset of
Fig. 2(c)) at the structural transition temperature was interpreted as the result of incipient magnetic
order.\cite{Fang} On doping 1\% Co, the anomaly temperature $T_{anom}$ was suppressed to 135 K. For 0.025 $<x<$
0.125, the resistivity anomaly disappears, instead, it shows a resistivity minimum at $T_{min}$ depending on the
Co-doping levels. At lower temperatures, these samples become superconducting with $T_c$ from 7 to 13 K, as can
be seen more clearly in Fig. 2(b). The samples of $x$=0.15 and 0.2 show no sign of superconducting transition
above 3 K. The magnetic susceptibility at low temperatures (Fig. 2(c)) show strong diamagnetic signal for the
samples with 0.025 $\leq x\leq $ 0.125. The magnetic expelling (Meissner effect) fraction and magnetic shielding
one of the sample of $x$=0.075 are estimated to be 11\% and 30\%, respectively, confirming bulk
superconductivity. For $x$=0, 0.01 and 0.15, no superconductivity was observed above 2 K.

\begin{figure}
\includegraphics[width=7cm]{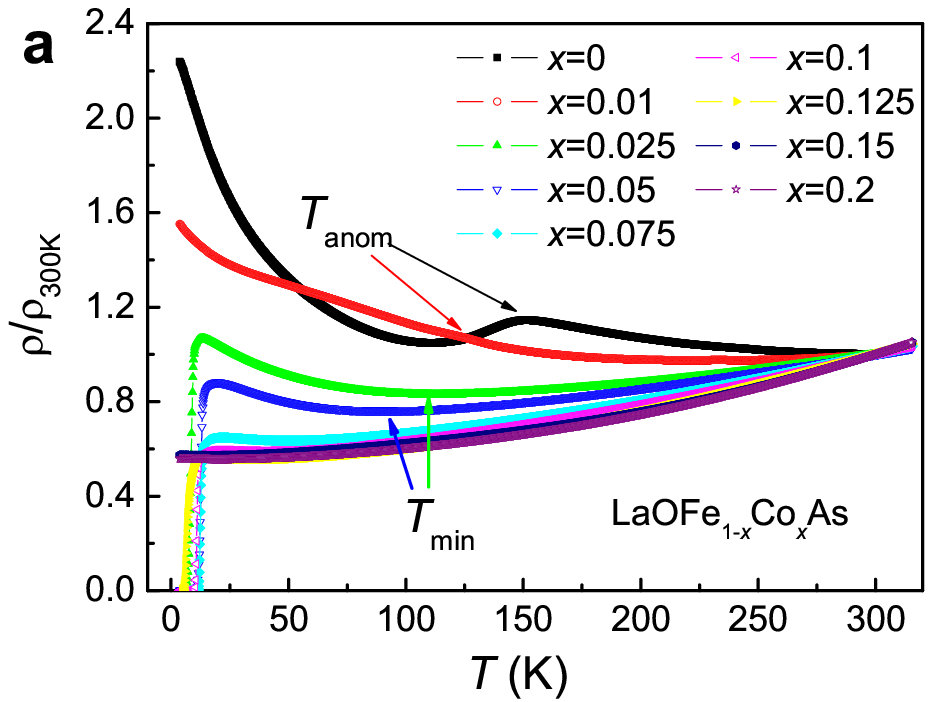}
\includegraphics[width=7cm]{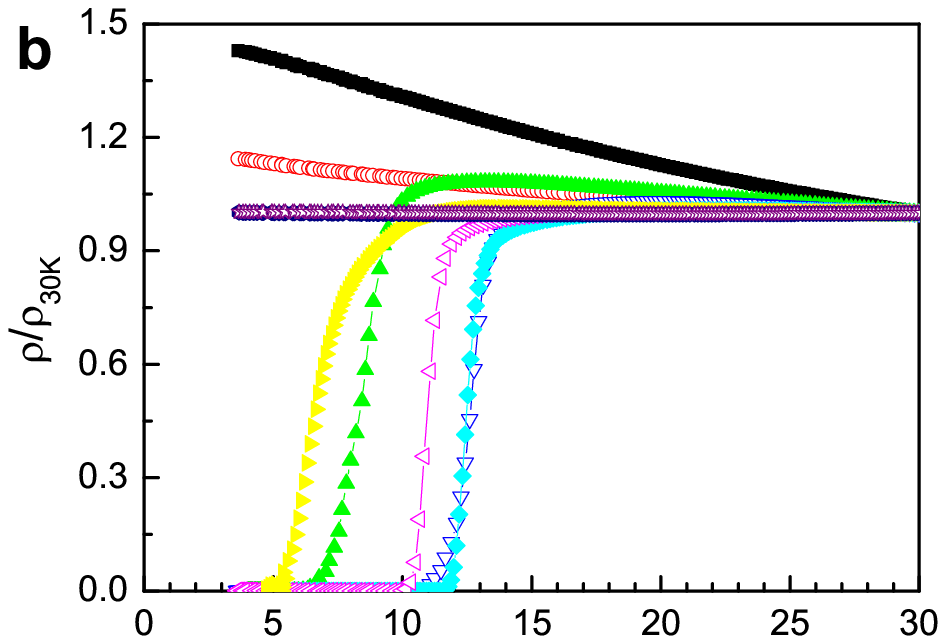}
\includegraphics[width=7cm]{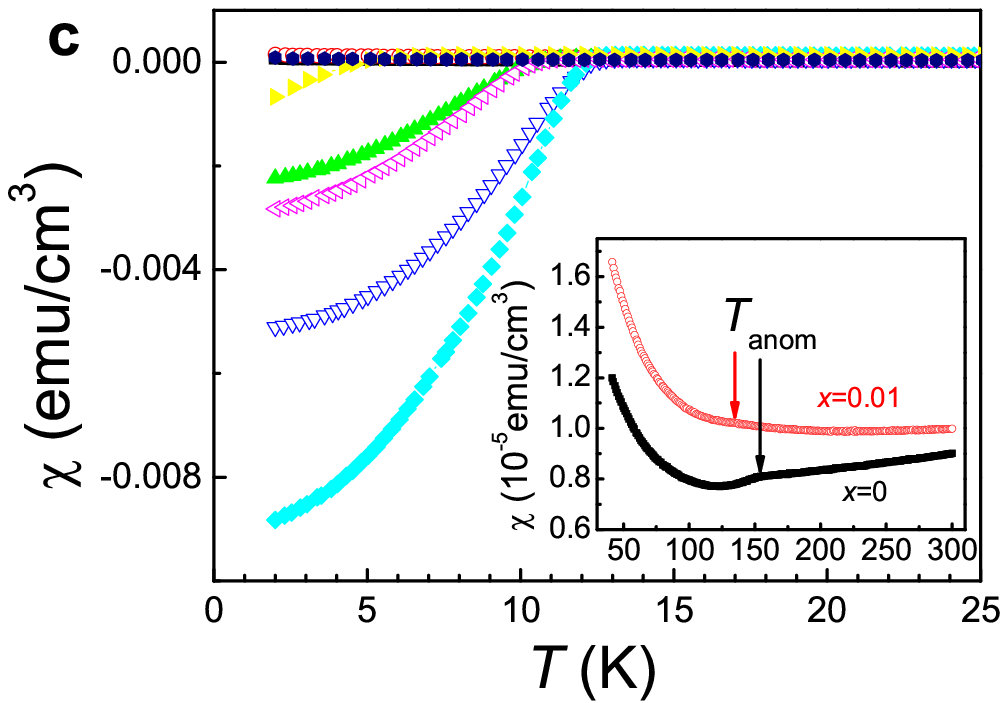}
\caption {\textbf{Electrical resistivity ($\rho$) and magnetic susceptibility ($\chi$) of
LaFe$_{1-x}$Co$_{x}$AsO samples.} \textbf{a}, $\rho(T)$ curves of LaFe$_{1-x}$Co$_{x}$AsO samples. The
resistivity was measured via a standard four-terminal method. The data are normalized to $\rho_{300K}$ because
the resistivity measured on polycrystalline samples is often higher than the intrinsic value due to the grain
boundary and surface effect. Nevertheless, it is here noted that the room temperature resistivity tends to
decrease with Co doping. \textbf{b}, $\rho(T)$ of LaFe$_{1-x}$Co$_{x}$AsO samples below 30 K, showing the
superconducting transitions. Normalized $\rho$ is again employed for clarity. \textbf{c}, $\chi(T)$ of
LaFe$_{1-x}$Co$_{x}$AsO samples, measured on a Quantum Design Magnetic Property Measurement System (MPMS-5).
Field-cooling protocols were used under the field of 10 Oe. The inset shows the $\chi(T)$ data for the samples
of $x$=0 and 0.01, measured under the magnetic field of 1000 Oe. A drop/kink in $\chi$ can be found at 150 K and
135 K for $x$=0 and 0.01, respectively.}
\end{figure}

The electronic phase diagram for LaFe$_{1-x}$Co$_{x}$AsO was thus established from the above experimental data,
as depicted in Figure 3. The phase region of the SDW state is very narrow. 2.5\% Co doping completely destroys
the SDW order, and superconductivity emerges. In the superconducting regime with 0.025 $\leq x\leq $ 0.125, one
sees a dome-like $T_c(x)$ curve, similar to that of cuprate superconductors. Though the normal state shows
metallic conduction at high temperatures, semiconducting behaviour is all observed above $T_c$. It is noted here
that the borderline between metallic and semiconducting regions is not well established because polycrystalline
samples were employed. For the higher Co-doping levels of $x\geq$0.15, superconductivity no longer survives.
Further Co doping is also of interest, because the other end member LaFeCoO was an itinerant ferromagnetic
metal\cite{LaCoAsO}.

\begin{figure}
\includegraphics[width=7cm]{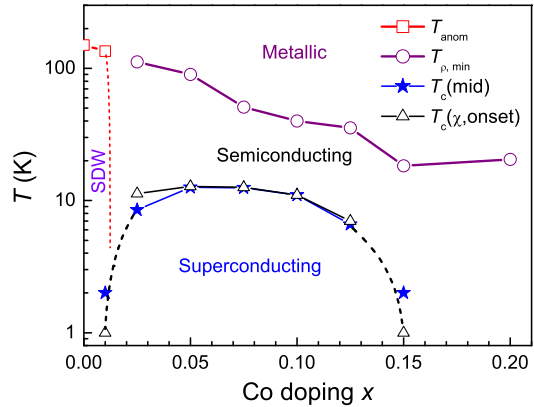}
\caption {\textbf{The electronic phase diagram for LaFe$_{1-x}$Co$_{x}$AsO.} $T_{anom}$ denotes the resistivity
anomaly temperature arising from the incipient magnetic order. $T_{\rho,min}$) separates the metallic and
semiconducting regions in the normal state of the superconductors. Note that the vertical axis is in logarithmic
scale.}
\end{figure}

The present Co-doped LaFeAsO system shows both similarities and differences in comparison with the phase diagram
of F-doped LaFeAsO\cite{Kamihara08,wnl,PD}. First, the AFM SDW state in LaFeAsO is suppressed or destroyed by
the electron doping in both systems. Second, both systems show a maximum $T_c$ upon electron doping. However,
there are some differences as for the details of the electronic phase diagrams. (1) Co doping destroys the AFM
order more strongly, and it shows no coexistence of superconductivity and SDW state. (2) The maximum $T_{c}$ is
significantly lower in Co-doped system. (3) The optimal doping level is significantly lower and the
superconducting region is narrower in LaFe$_{1-x}$Co$_{x}$AsO system. (4) The normal state of
LaFe$_{1-x}$Co$_{x}$AsO system shows semiconducting behaviour above $T_{c}$. The last three points are probably
related with the disorder effect within (Fe/Co)$_2$As$_2$ layers. The first issue can be qualitatively
understood in Figure 4. According to the theoretical studies\cite{Yildirim,Si,Lu}, the AFM order in the parent
compound originates from the competing nearest-neighbor and next-nearest-neighbor superexchange interactions,
bridged by As 4$p$ orbitals. Both interactions are antiferromagnetic, which gives rise to a frustrated magnetic
ground state (striped AFM). Upon doping Co onto the Fe site, the original antiferromagnetic superexchange
interactions may be changed into a double exchange between Co and Fe atoms, which obviously destroys the striped
AFM order.

\begin{figure}
\includegraphics[width=8cm]{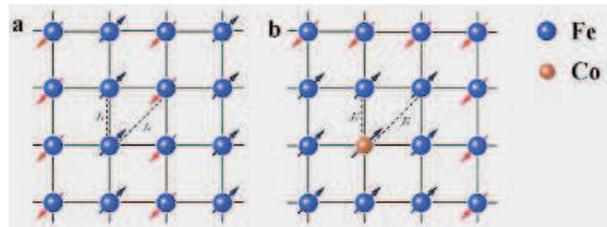}
\caption {\textbf{Destruction of AFM in Fe planes by Co doping}. \textbf{a}, the nearest-neighbor ($J_1$) and
next-nearest-neighbor ($J_2$) superexchange interactions result in a striped AFM order in Fe planes when $J_{2}
> 2J_{1} >0$. \textbf{b}, the adjacent interactions between Fe and Co become ferromagnetic ($J' < 0$) due to
double exchange, which strongly destroys the original frustrated AFM order.}
\end{figure}

Co-doping induced superconductivity challenges our previous understanding about occurrence of superconductivity
via chemical doping. As a typical magnetic element, cobalt does not act as superconducting Cooper pairs
breakers, which implies unconventional superconductivity. Additionally, superconductivity is robust in spite of
doping on the Fe$_2$As$_2$ conducting layers. These facts support the itinerant scenario of the 3$d$ electrons,
reminiscent of superconductivity on the border of itinerant-electron ferromagnetism in UGe$_2$\cite{Saxena}. For
the cuprate superconductors, in sharp contrast, substitution of Cu with its neighbors in Periodic Table (Ni and
Zn) in CuO$_2$ definitely destroys the superconductivity.\cite{Tarascon} Therefore, our result suggests that the
mechanisms of two class high temperature superconductivity should be essentially different.


\begin{acknowledgments}
This work is supported by the National Basic Research Program of China, the Natural Science Foundation of China
and the PCSIRT of the Ministry of Education of China.
\end{acknowledgments}


\begin{thebibliography}{00}
\bibitem{Bednorz&Muller}Bednorz, J. G. \& Muller, K. A. Possible high $T_c$ superconductivity in the Ba-La-Cu-O system. \emph{Z. Phys. B} \textbf{64}, 189-193 (1986).
\bibitem{Cava1988}Cava, R. J. \emph{et al.} Superconductivity near 30 K without copper: the Ba$_{0.6}$K$_{0.4}$BiO$_3$ perovskite. \emph{Nature} \textbf{332}, 814-816 (1988).
\bibitem{Hebard}Hebard, A. F. \emph{et al.} Superconductivity at 18 K in potassium-doped fullerene (C$_{60}$). \emph{Nature} \textbf{350}, 600-601 (1991).
\bibitem{Takada}Takada, K.\emph{ et al.} Superconductivity in two-dimensional CoO$_2$ layers. \emph{Nature} \textbf{422}, 53-55 (2003).
\bibitem{Kamihara08}Kamihara, Y. \emph{et al.} Iron-based layered superconductor La[O$_{1-x}$F$_x$]FeAs ($x$=0.05-0.12) with $T_c$=26 K. \emph{J. Am. Chem. Soc.} \textbf{130}, 3296-3297 (2008).
\bibitem{Dai}Cruz, C. \emph{et al.} Magnetic order versus superconductivity in the iron-based layered La(O$_{1-x}$F$_{x}$)FeAs systems. \emph{Nature} 453, 899-902 (2008).
\bibitem{Chen-Sm}Chen, X. H. \emph{et al.} Superconductivity at 43 K in samarium-arsenide oxides SmFeAsO$_{1-x}$F$_x$. \emph{Nature} \textbf{354}, 761-762 (2008).
\bibitem{Chen-Ce}Chen, G. F. \emph{et al.} Superconductivity at 41 K and its competition with spin-density-wave insability in layered CeO$_{1-x}$F$_x$FeAs. \emph{Phys. Rev. Lett.} 100, 247002 (2008).
\bibitem{Ren-Sm}Ren, Z. A. \emph{et al.} Superconductivity at 55 K in iron-based F-doped layered quaternary compound Sm[O$_{1-x}$F$_x$]FeAs. \emph{Chin. Phys. Lett.} \textbf{25}, 2215-2216 (2008).
\bibitem{Ren-Sm2}Ren, Z. A. \emph{et al.} Novel superconductivity and phase diagram in the iron-based arsenic-oxides ReFeAsO$_{1-\delta}$ (Re = rare-earth metal) without fluorine doping. \emph{Europhysics Lett.} \textbf{83}, 17002 (2008).
\bibitem{Wang}Wang, C. \emph{et al.} Thorium-doping induced superconductivity up to 56 K in Gd$_{1-x}$Th$_x$FeAsO. Preprint at http://arxiv.org/abs/0804.4290v2 (2008).
\bibitem{Cava}Cava, R. J. Oxide superconductors. \emph{J. Am. Ceram. Soc.} \textbf{83}, 5-28 (2000).
\bibitem{Singh}Singh, D. J. \& Du, M.-H. Density functional study of LaFeAsO$_{1-x}$F$_x$: A low carrier density superconductor near itinerant magnetism. \emph{Phys. Rev. Lett.} \textbf{100}, 237003 (2008).
\bibitem{wnl}Dong, J. \emph{et al.} Competing orders and spin-density-wave instability in La(O$_{1-x}$F$_x$)FeAs. \emph{Europhysics Lett.} \textbf{83}, 27006 (2008).
\bibitem{Tesanovic}Cvetkovic, V. \& Tesanovic, Z. Multiband magnetism and superconductivity in Fe-based compounds. Preprint at http://arxiv.org/abs/0804.4678 (2008).
\bibitem{Xu}Xu, G.\emph{ et al.} Doping-dependent phase diagram of LaOMAs (M=V-Cu) and electron-type superconductivity near ferromagnetic
instability. \emph{Europhysics Lett.} \textbf{82}, 67002 (2008).
\bibitem{Kito}Kito, H. Eisaki, H. \& Iyo, A. Superconductivity at 54 K in F-Free NdFeAsO$_{1-y}$. \emph{J. Phys. Soc. Jpn.} \textbf{77}, 063707
(2008).
\bibitem{Wu}Wu, G. et al. Superconductivity induced by oxygen deficiency in Sr-doped LaOFeAs. Preprint at http://arxiv.org/abs/0806.1687 (2008).
\bibitem{Fang}Fang, C. \emph{ et al.} Theory of electron nematic order in LaFeAsO. \emph{Phys. Rev. B.} \textbf{77}, 224509 (2008).
\bibitem{PD}Luetkens, H. \emph{et al.} Electronic phase diagram of the LaO$_{1-x}$F$_x$FeAs superconductor. Preprint at http://arxiv.org/abs/0806.3533 (2008).
\bibitem{LaCoAsO}Yanagi, H. \emph{et al.} Itinerant ferromagnetism in layered crystals LaCoOX (X = P, As). Preprint at http://arxiv.org/abs/0806.0123 (2008).
\bibitem{Yildirim}Yildirim, T. Origin of the $\sim$150 K anomaly in LaOFeAs: Competing antiferromagnetic superexchange interactions, frustration, and structural phase
transition. Preprint at http://arxiv.org/abs/0804.2252 (2008).
\bibitem{Si}Si, Q. \& Abrahams, E.  Strong correlations and magnetic frustration in the high $T_{c}$ iron
pnictides. Preprint at http://arxiv.org/abs/0804.2480 (2008).
\bibitem{Lu}Ma, F. Lu, Z. \& Xiang, T. Antiferromagnetic superexchange interactions in LaOFeAs. Preprint at http://arxiv.org/abs/0804.3370 (2008).
\bibitem{Saxena}Saxena, S. S.\emph{ et al.} Superconductivity on the border of itinerant-electron ferromagnetism in UGe$_2$. \emph{Nature} \textbf{406}, 587-592 (2000).
\bibitem{Tarascon}Tarascon, J. M. \emph{et al.} 3$d$-metal doping of the high-temperature superconducting perovskites La-Sr-Cu-O and
Y-Ba-Cu-O. \emph{Phys. Rev. B.} \textbf{36}, 8393-8400 (1987).

\end{thebibliography}
\end{document}